\documentclass[aps,pra,reprint,longbibliography,floatfix]{revtex4-2}
\usepackage{amsmath,amssymb,bm}
\usepackage{graphicx}
\usepackage{booktabs}

\usepackage{hyperref}
\usepackage[table]{xcolor}

\usepackage{tabularx}
\usepackage{ragged2e}
\newcolumntype{Y}{>{\RaggedRight\arraybackslash}X}

\newcommand{\ii}{\mathrm{i}}
\newcommand{\Pc}{\overline{P}_{c}}

\definecolor{tablehead}{RGB}{225,232,240}
\definecolor{tablerow}{RGB}{244,247,250}

\graphicspath{{Figures/}}
\begin{document}
\title{Drive-tunable Floquet exchange spectroscopy with Hong--Ou--Mandel interference}

\author{Eric~R.~Bittner}
\affiliation{Department of Physics, University of Houston, Houston, TX 77204, USA}
\email{ebittner@central.uh.edu}

\date{\today}
\begin{abstract}
Floquet driving reshapes cavity spectra, but single-photon sideband populations do not reveal how it changes two-photon exchange. Here we show theoretically that delay-resolved Hong--Ou--Mandel (HOM) coincidences can probe this response in a collectively coupled cavity. A frequency-entangled signal photon scatters linearly from the driven cavity while its idler provides a delayed reference. Averaging over an unsynchronized drive phase removes coherence between different net output orders but preserves exchange interference within each order. Fourier transformation of the coincidence trace yields two drive-dependent exchange branches associated with the Floquet-polariton spectrum. As the drive changes, the branches shift and exchange weight; the dominant physical output order also changes, a transfer not reproduced by single-photon sideband populations. We further map the coincidence-trace response over drive frequency and amplitude and connect it to a count-model-dependent Fisher information. Floquet control thus adds a tunable exchange spectrum to HOM interferometry and gives quantitative targets for an experiment.

\end{abstract}
\maketitle

\section{Introduction}
\label{sec:introduction}

Cavity quantum electrodynamics offers a way to control how light interacts
with matter. When many emitters share a cavity mode, their collective bright
excitation can couple strongly to light even when each emitter couples
weakly. This regime appears in molecular and rare-earth cavity systems and
motivates applications in spectroscopy and quantum interfaces
\cite{Shalabney2015MolecularResonators,Zhong2015RareEarthNanocavity}.
Periodic driving adds in-situ spectral control: the drive frequency sets
Floquet-sideband spacing, and the drive amplitude controls coupling between
sidebands. Here, we ask how this control reshapes the exchange response of
entangled photons.

Hong--Ou--Mandel (HOM) interference provides such a probe: a delay scan
measures how the two-photon coincidence rate changes when the detection
assignments are exchanged at a balanced beam splitter \cite{Hong1987HOM}.
Frequency-entangled photons have used this sensitivity to characterize
plasmonic and resonant media
\cite{Kalashnikov2014Plasmonic,Kalashnikov2017ResonantMedium}, and to retain
dynamical information absent from one-photon spectra
\cite{Dorfman2016QuantumLight,Schlawin2017EntangledSpectroscopy,
Mukamel2020Roadmap,Dorfman2021HOMSpectroscopy}. HOM metrology has also used
coincidence rates to estimate an unknown delay: Chen \emph{et al.} optimized
the Fisher information of a biphoton beat-note interferometer and demonstrated
fiber-delay sensing, while Scott \emph{et al.} showed how photon-number and
detection-time information can improve delay estimation and calibration
\cite{Chen2019HOMBeatNote,Scott2020BeyondCoincidence}. Here the delay is the
scan coordinate; the parameters of interest are the drive frequency and
amplitude that reshape the sample's full coincidence trace.

Floquet polaritons and driven cavity spectra have been engineered in several
platforms \cite{Clark2019InteractingFloquetPolaritons,
Baghdad2023SpectralPolaritons,Kuznetsov2026FloquetPolaritons}. A recent
proposal uses two modulated cavities to convert spectrally mismatched photons
into more indistinguishable frequency-comb states
\cite{Yu2026CavityFloquetHOM}. Our scheme uses one driven, collectively coupled
cavity as the sample and reads out its drive-dependent exchange spectrum. It
avoids coordinating the modulation phases of two cavities, but it does not
remove the need for strong collective coupling, fast resonance modulation,
and efficient photon collection. The two approaches target different tasks:
comb-state matching and Floquet spectroscopy, respectively.

Our earlier work used entangled-photon scattering and HOM
correlations to probe dynamical symmetry breaking in a collective ensemble
\cite{Li2018DynamicalSymmetry}; here, periodic driving makes the exchange
spectrum tunable. We have also developed an input/output scattering theory for
frequency-entangled biphotons interacting with cavity polaritons
\cite{piryatinski2026scattering}, studied topology and spectral entanglement
in cavity-mediated scattering \cite{bittner2025topology}, and formulated
biphoton spectroscopy using Lyapunov-based input--output dynamics
\cite{wgnx-l7qx}.

Here, a frequency-entangled pair probes a periodically driven cavity: the
signal photon scatters from the cavity, while the idler supplies a delayed
reference. In principle, this measurement applies to other linear Floquet
scatterers; collective coupling is not required. We use a collectively coupled
cavity because its driven light--matter polaritons provide a structured
spectrum whose motion appears in the exchange response. The cavity acts
linearly on the signal photon. The two-photon observable arises from exchange
interference at the final beam splitter, not from an interaction-generated
vertex. This distinguishes the present mechanism from our earlier studies of
exciton--exciton interactions and many-body photon entanglement
\cite{Bittner2020ExcitonInteractions,Li2019EntanglementEntropy}. When the
photon arrival time is not locked to the cavity modulation, averaging over
the drive phase removes coherence between different net Floquet output orders
while preserving interference among exchange pathways within each order.

For the collectively coupled cavity, we find two dominant delay-frequency
branches associated with the driven Floquet-polariton pair. Their frequencies
move with the drive and form a complementary pair over the parameter range
studied. These quantitative predictions use a single-mode, homogeneous
Tavis--Cummings model; disorder and additional cavity resonances may shift,
broaden, or add branches, and we leave those effects for future work. The
folded branch also
changes its dominant physical output order from $n=+1$ to $n=0$, a transfer not
reproduced by single-photon sideband populations. Finally, we construct a
response map over modulation frequency and amplitude from the full
coincidence trace. Prior HOM metrology optimizes estimates of delay; this map
instead characterizes how measured coincidence statistics respond to Floquet
controls. A count model can estimate precision for specified exposure and detector
assumptions; the plotted response metric instead ranks trace deformation
without count-rate weighting and can guide denser sampling of the drive
parameters.
The model, observable, and normalization are defined in
Sec.~\ref{sec:model}; Sec.~\ref{sec:spectroscopy} presents the exchange
spectrum; Sec.~\ref{sec:information} analyzes its coincidence-statistics
response; and Sec.~\ref{sec:discussion} discusses experimental prospects and
limits.

\section{Biphoton probe and collective cavity}
\label{sec:model}

\begin{figure}[t]
\includegraphics[width=\columnwidth]{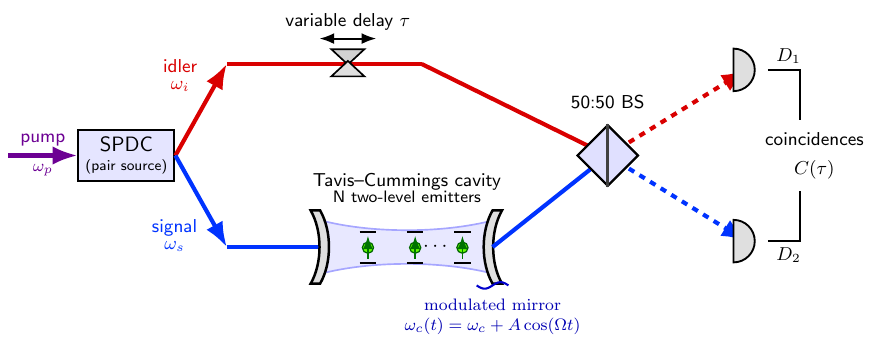}
\caption{Proposed HOM spectroscopy of a Floquet Tavis--Cummings
cavity. An SPDC source generates a frequency-entangled signal--idler
pair. The signal traverses a collectively coupled cavity--emitter
system with cavity frequency modulated as
$\omega_c(t)=\omega_c+A\sin(\Omega t)$, while the idler acquires a
controllable delay $\tau$. Coincidences after the balanced beam
splitter measure the exchange overlap of the outgoing photons.}
\label{fig:setup}
\end{figure}

We use Hong--Ou--Mandel (HOM) interference to probe how a periodically driven,
collectively coupled cavity changes a photon spectrum. One photon from an
entangled pair passes through the cavity; its partner bypasses the sample and
serves as a delayed reference. Their coincidence rate at a balanced beam
splitter measures the exchange overlap of the two photons. Sweeping the delay
and the cavity-drive frequency therefore reveals the drive-dependent
spectral response, even when the experiment does not resolve individual
Floquet sidebands. Figure~\ref{fig:setup} shows the arrangement.

We model the incident pair as
\begin{equation}
\lvert\Psi_{\mathrm{in}}\rangle=\iint d\omega_s d\omega_i\,
\Phi(\omega_s,\omega_i)
a_s^\dagger(\omega_s)a_i^\dagger(\omega_i)\lvert0\rangle,
\label{eq:inputstate}
\end{equation}
where $s$ and $i$ label the signal and idler. Write their physical frequencies
as $\widetilde\omega_s=\overline\omega_s+\omega_s$ and
$\widetilde\omega_i=\overline\omega_i+\omega_i$; $\omega_s$ and $\omega_i$
denote detunings from the carriers. The signal enters the cavity, while the
idler bypasses it and provides a delayed reference at the beam splitter. We
normalize the joint spectral amplitude (JSA) as
$\iint d\omega_s d\omega_i\lvert\Phi\rvert^2=1$ and choose
\begin{equation}
\Phi={\cal N}\exp\!\left[-\frac{(\omega_s+\omega_i)^2}{4\sigma_+^2}
-\frac{(\omega_s-\omega_i)^2}{4\sigma_-^2}\right],
\label{eq:jsa}
\end{equation}
where $\sigma_+$ and $\sigma_-$ are the rms intensity widths in the sum and
difference coordinates $\omega_+=\omega_s+\omega_i$ and
$\omega_- =\omega_s-\omega_i$. The pump sets $\sigma_+$; phase matching sets
the broader relative-frequency width $\sigma_-$. When $\sigma_+\ll\sigma_-$,
the photons are strongly frequency anticorrelated. Unless noted otherwise,
$\sigma_+=0.25\kappa$ and $\sigma_-=2\kappa$, so all spectral widths are quoted
relative to the cavity linewidth defined below.

\begin{figure*}[t]
\centering
\includegraphics[width=\textwidth]{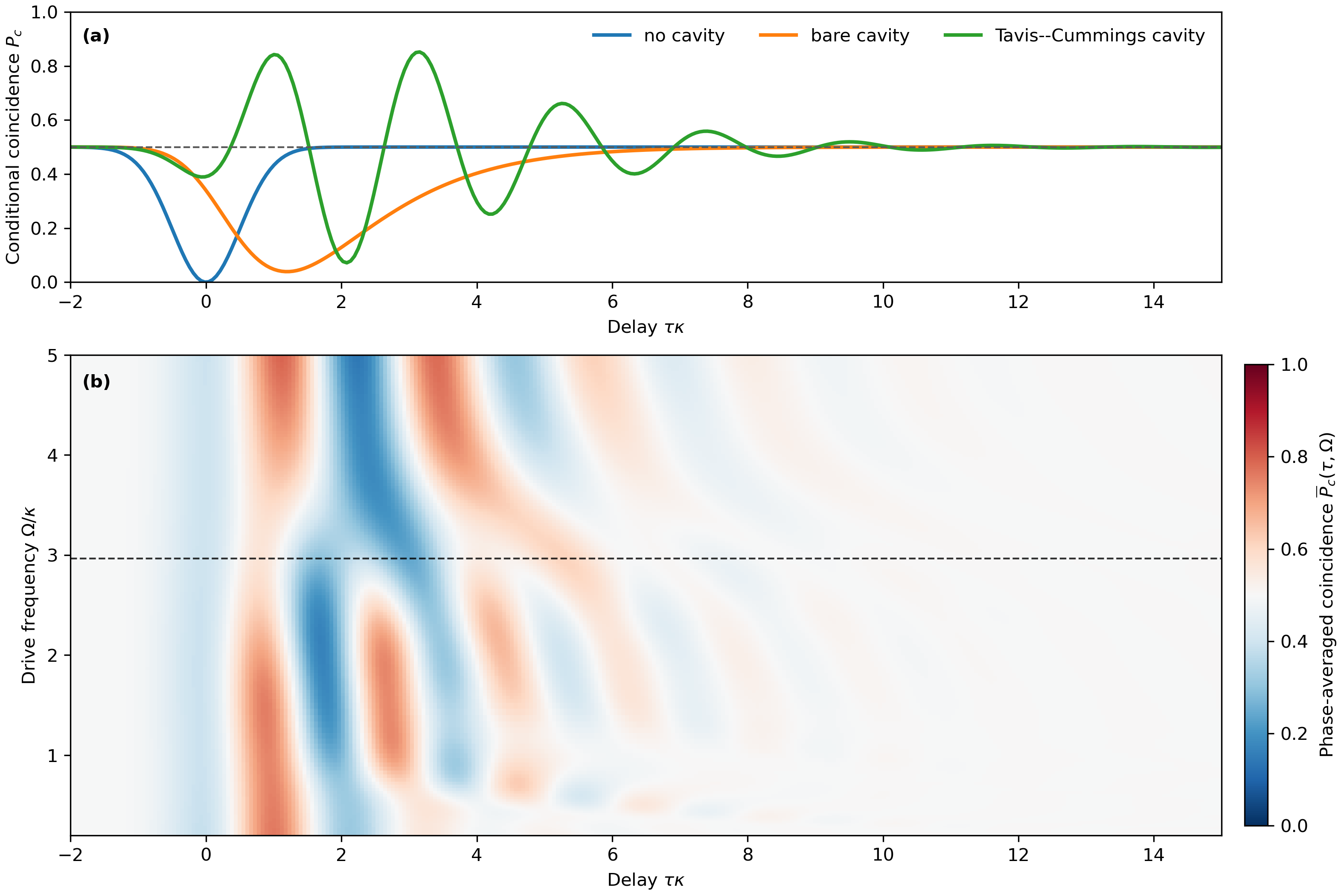}
\caption{Delay-domain and drive-frequency dependence of the same biphoton probe.
(a) Conditional HOM interferograms with no cavity, a bare cavity, and a static
Tavis--Cummings cavity ($A=0$). (b) Phase-averaged conditional coincidence
$\overline{P}_c(\tau,\Omega)$ for the Tavis--Cummings cavity as a function of
idler delay and positive drive frequency $0.2\leq\Omega/\kappa\leq5$. The
unmodulated $\Omega=0$ response is shown in (a) and is omitted from the lower
map. We use $A/\kappa=2.385$ and the drive convention
$A\sin(\Omega t)$. The color scale is the absolute coincidence probability.
The dashed line marks the static polariton splitting
$\Delta_{\mathrm{pol}}$ from Eq.~\eqref{eq:deltapol}.}
\label{fig:hom-baseline-comparison}
\end{figure*}

We denote the energy-decay rates through the input port, transmitted port,
and internal loss channels by $\kappa_1$, $\kappa_2$, and
$\kappa_{\mathrm{int}}$. Their sum
\begin{equation}
 \kappa=\kappa_1+\kappa_2+\kappa_{\mathrm{int}}
 \label{eq:kappadecomposition}
\end{equation}
is the total cavity intensity linewidth, so the intracavity field amplitude
decays at $\kappa/2$. Port 1 injects and reflects light; port 2 transmits it.
The collective material polarization has linewidth $\gamma$ and amplitude
decay rate $\gamma/2$. Here $\gamma$ represents homogeneous bright-mode
decoherence.

For $N$ identical emitters with lowering operators $\sigma_j^-$, the cavity
couples to the permutation-symmetric Dicke operator
\begin{equation}
 B=\frac{1}{\sqrt{N}}\sum_{j=1}^{N}\sigma_j^- .
 \label{eq:brightoperator}
\end{equation}
At excitation numbers much smaller than $N$, the Holstein--Primakoff expansion
gives $[B,B^\dagger]\simeq1$, so we treat $B$ as a bosonic mode. This
approximation requires weak excitation and negligible frequency disorder,
coupling variation, and bright--dark mixing. It fails at appreciable
excitation density or when inhomogeneous broadening and dark states matter.

The matrix element between the cavity and this bright excitation is enhanced
collectively to
\begin{equation}
 G=g\sqrt{N},
 \label{eq:collectivecoupling}
\end{equation}
where $g$ is the single-emitter vacuum coupling. For unequal couplings $g_j$,
the bright-mode strength is
$G=(\sum_j\lvert g_j\rvert^2)^{1/2}$ and the bright operator is weighted by
$g_j/G$.

The single-mode Tavis--Cummings reduction is a minimal description, not a
material-specific cavity model. Inhomogeneous broadening, bright--dark mixing,
and additional cavity resonances can shift, broaden, or add Floquet branches;
the predicted branch positions and weights are therefore model-specific.
Extending the scattering model to include these effects is a natural next step.

The Floquet equations use the cosine gauge
$\omega_c(t)=\omega_0+A\cos(\Omega t)$, which gives the couplings in
Eq.~\eqref{eq:floqueteq}. For $\Omega>0$, uniform averaging over the drive
phase makes this equivalent to the sine convention used in Figs.~\ref{fig:hom-baseline-comparison}
and \ref{fig:delay-spectrum}. The $\Omega=0$ reference is evaluated separately
with $A=0$, so it is the unmodulated cavity rather than a statically detuned
resonator. Thus $n$ labels frequency sidebands, not spatial modes. In a frame
rotating at $\omega_0$, the cavity and bright mode obey linear input--output
equations. An incident component at detuning
$\omega$ generates components at $\omega+n\Omega$. Eliminating the bright
mode gives the tridiagonal system below; the transmitted field is
$a_{\rm out,2}=-\sqrt{\kappa_2}a$.

Emitter-frequency modulation is a distinct scattering realization because the
ports couple to the cavity field, even though cavity and emitter modulation
are related through the relative instantaneous detuning of the closed system.
Its Floquet equations and exchange spectra are given in the Supplemental
Material.

Eliminating the collective polarization gives
\begin{align}
&\left[\frac{\kappa}{2}-\ii(\omega+n\Omega)
+\frac{G^2}{\gamma/2-\ii(\omega+n\Omega)}\right]a_n(\omega)
\nonumber\\
&\quad+\frac{\ii A}{2}\left[a_{n-1}(\omega)+a_{n+1}(\omega)\right]
=\sqrt{\kappa_1}\,\delta_{n0}.
\label{eq:floqueteq}
\end{align}
For unit incident amplitude, the transmitted and reflected Floquet amplitudes
are, up to the fixed port-phase convention,
\begin{equation}
 t_n(\omega)=-\sqrt{\kappa_2}a_n(\omega),\qquad
 r_n(\omega)=\delta_{n0}-\sqrt{\kappa_1}a_n(\omega).
 \label{eq:trfloquet}
\end{equation}
The results below take $t_n$ as the sample-arm transfer amplitude.
Equation~\eqref{eq:floqueteq} is solved as a tridiagonal system over
$-M\le n\le M$. It reduces to a modulated empty resonator when $G=0$
\cite{Minkov2017ModulatedResonator} and to the static collective transmission
when $A=0$.
For the drive-frequency maps in Figs.~\ref{fig:hom-baseline-comparison}
and \ref{fig:delay-spectrum}, the production cutoff is
$M(\Omega)=\min\{18,\max[5,\lceil A/\Omega\rceil+4]\}$; it ranges from
$5$ to $16$ over the plotted positive-drive grid. The response-geometry and
single-point phase-sensitive calculations use $M=5$.

\subsection{Phase-averaged HOM observable}
\label{sec:hom}

Let $\phi=\Omega t_0$ be the modulation phase when the signal arrives. At
fixed $\phi$, a process that shifts the signal by net Floquet order $n$
acquires phase $e^{\ii n\phi}$. The outgoing state is
\begin{equation}
 \lvert\Psi_{\mathrm{out}}(\phi)\rangle
 =\sum_n e^{\ii n\phi}\lvert\Psi_n\rangle ,
 \label{eq:phaseresolvedstate}
\end{equation}
where $\lvert\Psi_n\rangle$ sums all paths that end in the same net order.
The order labels the signal's net frequency shift, not a path through the
Floquet ladder; different absorption and emission sequences can interfere
within $\lvert\Psi_n\rangle$.

If photon arrival is unsynchronized, successive pairs sample $\phi$
uniformly. Averaging the density operator over $\phi$ gives
\begin{align}
 \overline{\rho}_{\mathrm{out}}
 &=\frac{1}{2\pi}\int_0^{2\pi}d\phi\,
   \lvert\Psi_{\mathrm{out}}(\phi)\rangle
   \langle\Psi_{\mathrm{out}}(\phi)\rvert \nonumber\\
 &=\sum_{n,m}
   \left[
   \frac{1}{2\pi}\int_0^{2\pi}d\phi\,
   e^{\ii(n-m)\phi}
   \right]
   \lvert\Psi_n\rangle\langle\Psi_m\rvert \nonumber\\
 &=\sum_n\lvert\Psi_n\rangle\langle\Psi_n\rvert .
 \label{eq:phaseaverage}
\end{align}
Phase averaging removes coherence between different net Floquet orders but
preserves interference among paths within each order. Floquet-order
populations, within-order pathway interference, and HOM exchange interference
are distinct quantities.

For net Floquet order $n$, an incident signal at detuning
$\omega_1-n\Omega$ exits the sample at $\omega_1$. The corresponding
biphoton amplitude at the two beam-splitter inputs is
\begin{equation}
 \Psi_n(\omega_1,\omega_2)=
 t_n(\omega_1-n\Omega)\Phi(\omega_1-n\Omega,\omega_2).
 \label{eq:output-amplitude}
\end{equation}
The order-resolved state is
\begin{equation}
 \lvert\Psi_n\rangle
 =
 \iint d\omega_1d\omega_2\,
 \Psi_n(\omega_1,\omega_2)
 a_1^\dagger(\omega_1)a_2^\dagger(\omega_2)
 \lvert0\rangle ,
 \label{eq:psin}
\end{equation}
where $\omega_1$ and $\omega_2$ enter the two ports of the final balanced beam
splitter. The total weight in order $n$ is
\begin{equation}
 W_n=
 \iint d\omega_1d\omega_2\,
 \left|\Psi_n(\omega_1,\omega_2)\right|^2 .
 \label{eq:wn}
\end{equation}
HOM interference depends on the overlap between each amplitude and its
frequency-exchanged counterpart. A relative delay $\tau$ in one arm gives
\begin{equation}
 X_n(\tau)=
 \iint d\omega_1d\omega_2\,
 \Psi_n(\omega_1,\omega_2)
 \Psi_n^*(\omega_2,\omega_1)
 e^{-\ii(\omega_1-\omega_2)\tau}.
 \label{eq:xn}
\end{equation}
The integrand pairs the two indistinguishable detector assignments. The delay
phase $e^{-\ii(\omega_1-\omega_2)\tau}$ is conjugate to their frequency
difference.

Frequency-insensitive coincidence detection after the balanced beam splitter
gives the phase-averaged probability
\begin{equation}
 \boxed{
 \Pc(\tau)=\frac12\left[
 1-\frac{\operatorname{Re}\sum_nX_n(\tau)}
         {\sum_nW_n}
 \right].
 }
 \label{eq:homprob}
\end{equation}
The denominator conditions on transmission, so $\Pc$ describes detected-pair
exchange statistics, not absolute pair transmission. Loss and Floquet-order
redistribution enter through $\Psi_n$ and $W_n$; the normalized trace measures
their effect on two-photon indistinguishability.

Define the complex normalized exchange amplitude
\begin{equation}
 Z(\tau)
 =\frac{\sum_nX_n(\tau)}{\sum_nW_n},
 \qquad
 \Pc(\tau)=\frac12[1-\operatorname{Re}Z(\tau)].
 \label{eq:complexexchange}
\end{equation}
The measured probability depends on $\operatorname{Re}Z$, while the full
complex amplitude also tracks phase rotation. Positive $\operatorname{Re}Z$
suppresses coincidences; negative $\operatorname{Re}Z$ raises them above the
distinguishable-photon value $1/2$. This enhancement can result from coherent
phase rotation, not only from a change in indistinguishability. The
Supplemental Material gives the instrument-response convolution and
order-resolved amplitudes.

\subsection{Parameterization of the model}
\label{sec:parameterization}

Unless stated otherwise, we use a lossless symmetric two-sided cavity with
$\kappa_1=\kappa_2=\kappa/2$ and $\kappa_{\mathrm{int}}=0$, and set
$G=1.5\kappa$ and $\gamma=0.1\kappa$. Numerically, $\kappa=1$; frequencies
are in units of $\kappa$ and delays in units of $\kappa^{-1}$.
The biphoton bandwidths are $\sigma_+=0.25\kappa$ and
$\sigma_-=2.0\kappa$.

We evaluate shifted frequencies, JSA values, and transmission amplitudes
directly. We check convergence against frequency resolution and Floquet
cutoff, including power in the edge orders $\lvert n\rvert=M$.

For zero modulation the resonant pole splitting is
\begin{equation}
\Delta_{\mathrm{pol}}
=2\sqrt{G^2-\frac{(\kappa-\gamma)^2}{16}}
=2.966\kappa .
\label{eq:deltapol}
\end{equation}
Thus $2G=3\kappa$ is a strong-coupling reference for these parameters, not a
general formula for the polariton splitting.

We scan drive frequencies and amplitudes over a few cavity linewidths. The
exchange-spectrum maps use $A/\kappa=2.385$, where the response scan shows
strong joint sensitivity to frequency and amplitude. The Supplemental
Material gives the response metric and selection criterion. This point lies
in the sampled control space; it is not another cavity resonance. The
Supplemental Material also converts these dimensionless parameters to
representative device scales.


\section{Drive-tunable Floquet exchange spectroscopy}
\label{sec:spectroscopy}

For each net Floquet output order $n$, the HOM exchange amplitude contains
the delay phase
\begin{equation}
X_n(\tau)=
\iint d\omega_1d\omega_2\,
\Psi_n(\omega_1,\omega_2)
\Psi_n^*(\omega_2,\omega_1)
e^{-\ii(\omega_1-\omega_2)\tau}.
\label{eq:exchange-delay}
\end{equation}
Fourier transformation of the coincidence trace resolves frequency
differences selected by exchange interference, not optical output frequencies
or Floquet quasienergies alone. We define the finite-aperture delay-frequency spectrum as
\begin{equation}
\widetilde{\delta P}_{11}(\nu;\Omega)
=
\int_{-2/\kappa}^{15/\kappa}d\tau\,
\left[\Pc(\tau;\Omega)-\frac12\right]e^{-\ii\nu\tau}.
\label{eq:delayft}
\end{equation}
The transform uses a rectangular delay window; its duration sets an angular
frequency resolution of about $0.37\kappa$. The displayed frequency grid is
more finely sampled, and does not increase that physical resolution. The
Supplemental Material tests window and sampling effects. Figure~\ref{fig:hom-baseline-comparison}(b)
shows the underlying coincidence map at $A/\kappa=2.385$ with drive
$A\sin(\Omega t)$. Its unmodulated trace is shown in
Fig.~\ref{fig:hom-baseline-comparison}(a), so the map omits $\Omega=0$.

The two dominant dispersive features are consistent with exchange between
Floquet replicas of the driven polariton pair. Let $\delta_F(\Omega)$ be the
same-zone quasienergy gap of the tracked pair. The Supplemental Material
derives the two dominant exchange-beat assignments and defines
$\Delta_F(\Omega)=2\Omega-\delta_F(\Omega)$ for the extended branch. The
observed frequencies then obey
\begin{align}
\nu_{\rm ext}&=\Delta_F(\Omega),\qquad
\nu_{\rm fold}=2\Omega-\Delta_F(\Omega),\\
\nu_{\rm ext}+\nu_{\rm fold}&=2\Omega.
\label{eq:branch-complementarity}
\end{align}
The relation identifies complementary exchange pathways in the parameter
window studied; it is not a universal HOM selection rule. The detailed
quasienergy-replica derivation and alternative beat assignments are given in
the Supplemental Material.

\begin{figure*}[!t]
\includegraphics[width=\textwidth]{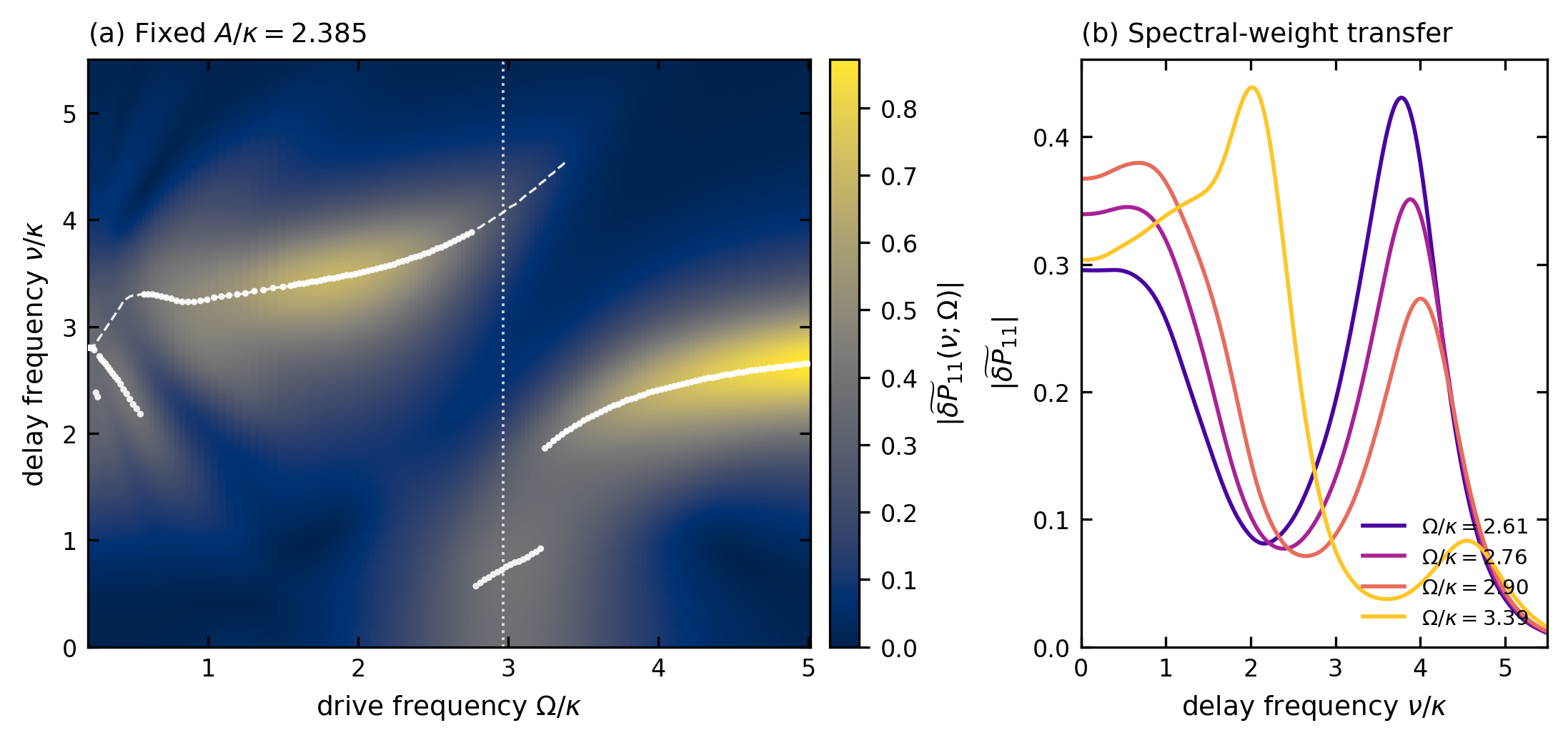}
\caption{Drive-tunable Floquet exchange spectrum at fixed
$A/\kappa=2.385$. (a) Magnitude of the Fourier transform of
$\Pc(\tau)-1/2$ versus drive frequency $\Omega$ and exchange-beat frequency
$\nu$. White markers track the two dominant complementary branches; the
vertical dotted line marks the static polariton splitting
$\Delta_{\rm pol}$. (b) Selected spectra show the drive-dependent transfer
of exchange weight between the branches.}
\label{fig:delay-spectrum}
\end{figure*}

Figure~\ref{fig:delay-spectrum} shows the central result: changing $\Omega$
moves both exchange branches and changes their relative HOM weight. Their
positions follow the dressed Floquet-polariton pair, while their visibility
depends on how that pair enters the exchanged two-photon amplitudes.

The output order $n$ records the signal photon's net frequency shift; it is
distinct from a quasienergy replica and an exchange-beat frequency. The
Supplemental Material shows that the folded branch persists while its
dominant output-order contribution changes from $n=+1$ to $n=0$. The extended
branch remains mainly in $n=0$, and single-photon sideband populations do not
show the same inversion. Sideband populations measure $W_n$; HOM interference
measures the permutation-sensitive amplitudes $X_n(\tau)$ in
Eq.~\eqref{eq:xn}. The drive can therefore tune branch frequencies and their
exchange-carrying output order separately.

The Supplemental Material gives phase-sensitive controls, complex channel
amplitudes, window-robustness checks, and transfer trajectories. It also
analyzes emitter-frequency modulation, which produces a distinct scattering
response. Here we focus on cavity-frequency modulation.


\section{Coincidence statistics and response geometry}
\label{sec:information}

HOM metrology has used coincidence rates and detector-time records to estimate
relative delay \cite{Chen2019HOMBeatNote,Scott2020BeyondCoincidence}. Here delay
is the measured scan coordinate; drive frequency $\Omega$ and amplitude $A$
are the controls. The response map
\begin{equation}
(\Omega,A)\longmapsto \Pc(\tau;\Omega,A)
\label{eq:response-map}
\end{equation}
predicts how the full delay-resolved coincidence probability changes across
Floquet controls. An experiment would reconstruct this map by scanning delay
and recording coincidences at each drive setting.

Because $\Pc$ is conditioned on transmission through the cavity, incident-pair
exposure $\mathcal N_i$ in delay bin $\tau_i$ gives the Poisson mean
\begin{equation}
\Lambda_i=\mathcal N_i\,\mathcal T_s(\Omega,A)\,
\Pc(\tau_i;\Omega,A),
\label{eq:poisson-model}
\end{equation}
where $\mathcal T_s$ is the spectrum-averaged signal transmission. The
count-model Fisher matrix is
\begin{equation}
\begin{aligned}
\mathcal I_{\mu\nu}&=\sum_i
\frac{\mathcal N_i}{\mathcal T_s\Pc(\tau_i)}\\
&\quad\times\partial_\mu[\mathcal T_s\Pc(\tau_i)]
\partial_\nu[\mathcal T_s\Pc(\tau_i)].
\end{aligned}
\label{eq:fisher-HOM}
\end{equation}
Here $\lambda^\mu\in\{\Omega,A\}$ indexes the two controls.
It includes throughput and conditional trace changes; a statistical sensitivity
claim would require a specified exposure, detector model, and noise budget.

\begin{figure*}[!t]
\centering
\includegraphics[width=\textwidth]{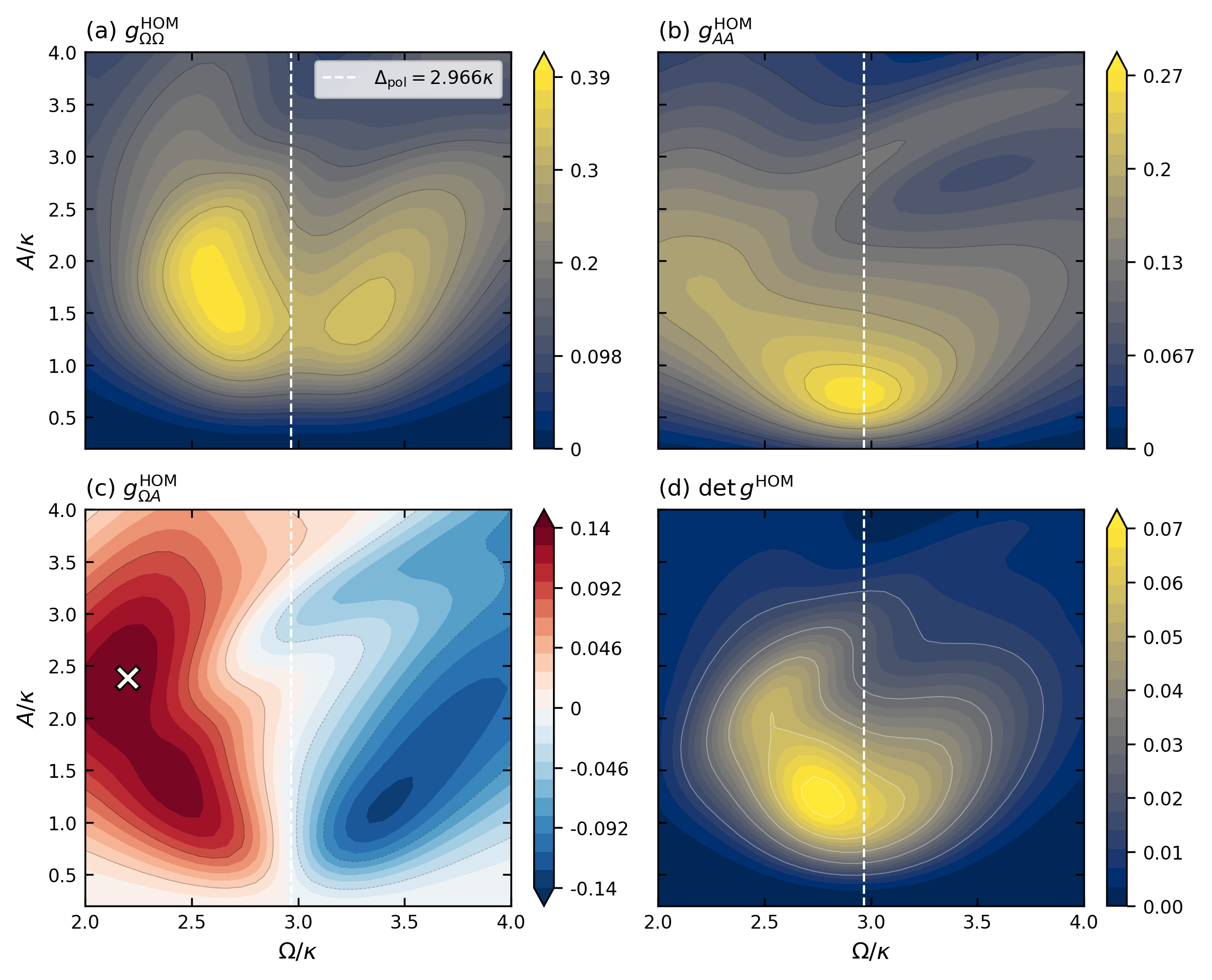}
\caption{Predicted HOM response metric over drive controls $(\Omega,A)$. The
diagonal components measure the integrated trace response to each control;
$g_{\Omega A}$ measures their alignment, and $\det g$ the squared area spanned
by the two local responses. The cross in panel (c) marks
$(\Omega,A)/\kappa=(2.2,2.385)$, selected for the phase-sensitive controls in
the Supplemental Material.}
\label{fig:metric}
\end{figure*}

Figure~\ref{fig:metric} shows a complementary, unweighted response metric,
\begin{equation}
 g^{\mathrm{HOM}}_{\mu\nu}=\int d\tau\,
 \partial_\mu\Pc(\tau)\,\partial_\nu\Pc(\tau),
 \qquad \lambda^\mu\in\{\Omega,A\}.
 \label{eq:metric}
\end{equation}
It maps how neighboring conditional coincidence traces deform over the
chosen delay interval, independent of count rate; it is not a Fisher-information
bound or a precision estimate. The mixed component measures alignment of the
two local trace changes, while
$\det g=g_{\Omega\Omega}g_{AA}-g_{\Omega A}^2$ is their squared spanned
area. An experiment can use the diagonal components to identify drive regions
where small changes in either control most strongly reshape the trace, then
sample those regions more densely. Whether finite data can distinguish the settings depends on the
count rate, exposure, and detector model.

The selected point maximizes $g_{\Omega A}$ on the sampled grid, so the two
control changes produce strongly aligned local trace deformations there; it
does not maximize statistical information. Near the avoided crossing, Floquet
hybridization redistributes exchange pathways, and small control changes
reshape the trace. The response metric can therefore guide where to refine a
scan of $(\Omega,A)$. The Fisher matrix determines whether the resulting
changes are statistically resolvable for a specified throughput, exposure,
and detector model.


\section{Discussion and conclusion}
\label{sec:discussion}

The central result is spectral tunability. Changing the drive moves the
Floquet--HOM exchange branches and redistributes their weight; the folded
branch also changes its dominant physical output order. Ordinary HOM delay
scans measure one interferogram at a time. Here the driven cavity adds
frequency and amplitude controls, allowing the full coincidence response to
be programmed while the input biphoton stays fixed. This is the added
capability beyond preparing a new light--matter state.

The approach complements the two-cavity Floquet proposal for matching
spectrally mismatched photons \cite{Yu2026CavityFloquetHOM}. Using one driven
sample avoids controlling the relative modulation phase of two cavities, but
it does not by itself establish a more feasible device. At the principal
operating point, the model requires $\Omega/(2\pi)=70.8\,\mathrm{GHz}$ and a
peak cavity-frequency excursion $A/(2\pi)=76.8\,\mathrm{GHz}$. It predicts
spectrum-averaged signal transmission of $15.2\%$, compared with $23.3\%$ for
the static cavity. These scales make modulation depth, useful port coupling,
and coincidence collection central design requirements. Integrated
electro-optic systems have reached relevant drive frequencies
\cite{Li2020LNModulator,Multani2026MillimeterWaveEO}, but the cited work does
not combine this excursion with a collectively coupled narrow-line medium and
efficient two-photon collection. The theory specifies a target, not a
realized device or a sensing advantage.

The prediction assumes weak excitation, a homogeneous collective bright
mode, one cavity resonance, and frequency-insensitive coincidence detection.
The cavity acts linearly on one photon; the two-photon response comes from
exchange interference at the final beam splitter, with no saturation or
connected two-photon vertex. The measured coincidence probability accesses
only the real part of the normalized exchange amplitude. Finite drive-phase
synchronization, detector jitter, internal loss, and material disorder will
set the observed contrast and frequency resolution. The Supplemental Material
analyzes instrumental smoothing and emitter-frequency modulation, which
provides a distinct fixed-input scattering response and a potential route for
strain or Stark actuation.

These results establish a theoretical route to Floquet exchange spectroscopy:
use one driven collective cavity to tune the frequencies and output-channel
character of a biphoton coincidence spectrum, then infer that response from
delay-resolved counts. The response metric can guide where to refine the scan
of drive frequency and amplitude by locating regions where small changes in
either control most strongly reshape the conditional coincidence trace. It does not predict
parameter precision by itself; that requires a count model with specified
throughput, exposure, and detector statistics.

\section*{Author Contribution}

E.R.B. conceived the project, developed the theoretical framework, performed the analytical and numerical calculations, analyzed and interpreted the results, and wrote the manuscript.
The author thanks Prof. Ajay Ram Srimath Kandada (Wake Forest)
 for useful discussions and comments on the work.

\section*{Data Availability}
The data supporting the findings of this study are publicly available on the \url{https://github.com/ebitnet65/Floquet_Tavis_Cummings}.

\begin{acknowledgments}
The work at the University of Houston was supported by the National Science Foundation under CHE-2404788 and the Robert A. Welch Foundation (E-1337).
\end{acknowledgments}

\bibliography{notes/floquet_tavis_cummings}

@article{Hong1987HOM,
  author  = {Hong, C. K. and Ou, Z. Y. and Mandel, L.},
  title   = {Measurement of subpicosecond time intervals between two photons by interference},
  journal = {Physical Review Letters},
  volume  = {59},
  number  = {18},
  pages   = {2044--2046},
  year    = {1987},
  doi     = {10.1103/PhysRevLett.59.2044}
}

@article{Kalashnikov2014Plasmonic,
  author  = {Kalashnikov, Dmitry A. and Pan, Zhenying and Kuznetsov, Arseniy I. and Krivitsky, Leonid A.},
  title   = {Quantum spectroscopy of plasmonic nanostructures},
  journal = {Physical Review X},
  volume  = {4},
  number  = {1},
  pages   = {011049},
  year    = {2014},
  doi     = {10.1103/PhysRevX.4.011049}
}

@article{Dorfman2016QuantumLight,
  author  = {Dorfman, Konstantin E. and Schlawin, Frank and Mukamel, Shaul},
  title   = {Nonlinear optical signals and spectroscopy with quantum light},
  journal = {Reviews of Modern Physics},
  volume  = {88},
  number  = {4},
  pages   = {045008},
  year    = {2016},
  doi     = {10.1103/RevModPhys.88.045008}
}

@article{Schlawin2017EntangledSpectroscopy,
  author  = {Schlawin, Frank},
  title   = {Entangled photon spectroscopy},
  journal = {Journal of Physics B: Atomic, Molecular and Optical Physics},
  volume  = {50},
  number  = {20},
  pages   = {203001},
  year    = {2017},
  doi     = {10.1088/1361-6455/aa8a7a}
}

@article{Mukamel2020Roadmap,
  author  = {Mukamel, Shaul and Freyberger, Matthias and Schleich, Wolfgang and Bellini, Marco and Zavatta, Alessandro and Leuchs, Gerd and Silberhorn, Christine and Boyd, Robert W. and S\'{a}nchez-Soto, Luis Lorenzo and Stefanov, Andr\'{e} and Barbieri, Marco and Paterova, Anna and Krivitsky, Leonid and Shwartz, Sharon and Tamasaku, Kenji and Dorfman, Konstantin and Schlawin, Frank and Sandoghdar, Vahid and Raymer, Michael and Marcus, Andrew and Varnavski, Oleg and Goodson, Theodore and Zhou, Zhi-Yuan and Shi, Bao-Sen and Asban, Shahaf and Scully, Marlan and Agarwal, Girish and Peng, Tao and Sokolov, Alexei V. and Zhang, Zhe-Dong and Zubairy, M. Suhail and Vartanyants, Ivan A. and del Valle, Elena and Laussy, Fabrice},
  title   = {Roadmap on quantum light spectroscopy},
  journal = {Journal of Physics B: Atomic, Molecular and Optical Physics},
  volume  = {53},
  number  = {7},
  pages   = {072002},
  year    = {2020},
  doi     = {10.1088/1361-6455/ab69a8}
}

@article{Dorfman2021HOMSpectroscopy,
  author  = {Dorfman, Konstantin E. and Asban, Shahaf and Gu, Bing and Mukamel, Shaul},
  title   = {{Hong--Ou--Mandel} interferometry and spectroscopy using entangled photons},
  journal = {Communications Physics},
  volume  = {4},
  number  = {1},
  pages   = {49},
  year    = {2021},
  doi     = {10.1038/s42005-021-00542-2}
}

@article{Chen2019HOMBeatNote,
  author  = {Chen, Yuanyuan and Fink, Matthias and Steinlechner, Fabian and Torres, Juan P. and Ursin, Rupert},
  title   = {{Hong--Ou--Mandel} interferometry on a biphoton beat note},
  journal = {npj Quantum Information},
  volume  = {5},
  number  = {1},
  pages   = {43},
  year    = {2019},
  doi     = {10.1038/s41534-019-0161-z}
}

@article{Scott2020BeyondCoincidence,
  author  = {Scott, Hannah and Branford, Dominic and Westerberg, Niclas and Leach, Jonathan and Gauger, Erik M.},
  title   = {Beyond coincidence in {Hong--Ou--Mandel} interferometry},
  journal = {Physical Review A},
  volume  = {102},
  number  = {3},
  pages   = {033714},
  year    = {2020},
  doi     = {10.1103/PhysRevA.102.033714}
}

@article{bittner2025topology,
  author  = {Bittner, Eric R. and Piryatinski, Andrei},
  title   = {Topology and spectral entanglement in cavity-mediated photon scattering},
  journal = {The Journal of Chemical Physics},
  volume  = {163},
  number  = {22},
  pages   = {224109},
  year    = {2025},
  doi     = {10.1063/5.0303703}
}

@article{piryatinski2026scattering,
  author  = {Piryatinski, Andrei and Jacobus, Nishaant and Dambal, Sameer and Bittner, Eric R. and Zhang, Yu and Kandada, Ajay Ram Srimath},
  title   = {Scattering theory of frequency-entangled biphoton states facilitated by cavity polaritons},
  journal = {Physical Review A},
  volume  = {113},
  number  = {3},
  pages   = {033730},
  year    = {2026},
  doi     = {10.1103/2kfb-ny9k}
}

@article{wgnx-l7qx,
  author  = {Dambal, Sameer and Kandada, Ajay Ram Srimath and Bittner, Eric R.},
  title   = {Quantum spectroscopy with biphotons: {Lyapunov}-based input-output dynamics},
  journal = {Physical Review A},
  volume  = {112},
  number  = {5},
  pages   = {053715},
  year    = {2025},
  month   = nov,
  doi     = {10.1103/wgnx-l7qx}
}

@article{Yu2026CavityFloquetHOM,
  author  = {Yu, Jia-Wang and Zhou, Xiao-Qing and Ni, Zhi-Bo and Cheng, Xiao-Tian and Zhao, Yi and Zhu, Hui-Hui and Li, Chen-Hui and Liu, Feng and Jin, Chao-Yuan},
  title   = {Enhancing Photon Indistinguishability of Spectrally Mismatched Single Photons by Cavity {Floquet} Engineering},
  journal = {Physical Review Letters},
  volume  = {136},
  number  = {5},
  pages   = {053601},
  year    = {2026},
  doi     = {10.1103/ww5v-ct3q},
  eprint  = {2507.02460},
  archivePrefix = {arXiv}
}

@article{Clark2019InteractingFloquetPolaritons,
  author  = {Clark, Logan W. and Jia, Ningyuan and Schine, Nathan and Baum, Claire and Georgakopoulos, Alexandros and Simon, Jonathan},
  title   = {Interacting {Floquet} polaritons},
  journal = {Nature},
  volume  = {571},
  number  = {7766},
  pages   = {532--536},
  year    = {2019},
  doi     = {10.1038/s41586-019-1354-5}
}

@article{Baghdad2023SpectralPolaritons,
  author  = {Baghdad, Mohamed and Bourdel, Pierre-Antoine and Schwartz, Sylvain and Ferri, Francesco and Reichel, Jakob and Long, Romain},
  title   = {Spectral engineering of cavity-protected polaritons in an atomic ensemble},
  journal = {Nature Physics},
  volume  = {19},
  number  = {8},
  pages   = {1104--1109},
  year    = {2023},
  doi     = {10.1038/s41567-023-02035-1}
}

@article{Minkov2017ModulatedResonator,
  author  = {Minkov, Momchil and Shi, Yu and Fan, Shanhui},
  title   = {Exact solution to the steady-state dynamics of a periodically modulated resonator},
  journal = {APL Photonics},
  volume  = {2},
  number  = {7},
  pages   = {076101},
  year    = {2017},
  doi     = {10.1063/1.4985381}
}

@article{Kalashnikov2017ResonantMedium,
  author  = {Kalashnikov, Dmitry A. and Melik-Gaykazyan, Elizaveta V. and Kalachev, Alexey A. and Yu, Ye Feng and Kuznetsov, Arseniy I. and Krivitsky, Leonid A.},
  title   = {Quantum interference in the presence of a resonant medium},
  journal = {Scientific Reports},
  volume  = {7},
  pages   = {11444},
  year    = {2017},
  doi     = {10.1038/s41598-017-11694-z}
}

@article{Li2018DynamicalSymmetry,
  author  = {Li, Hao and Piryatinski, Andrei and Jerke, Jonathan and Kandada, Ajay Ram Srimath and Silva, Carlos and Bittner, Eric R.},
  title   = {Probing dynamical symmetry breaking using quantum-entangled photons},
  journal = {Quantum Science and Technology},
  volume  = {3},
  number  = {1},
  pages   = {015003},
  year    = {2018},
  doi     = {10.1088/2058-9565/aa93b6}
}

@article{Li2019EntanglementEntropy,
  author  = {Li, Hao and Piryatinski, Andrei and Srimath Kandada, Ajay Ram and Silva, Carlos and Bittner, Eric R.},
  title   = {Photon entanglement entropy as a probe of many-body correlations and fluctuations},
  journal = {The Journal of Chemical Physics},
  volume  = {150},
  number  = {18},
  pages   = {184106},
  year    = {2019},
  doi     = {10.1063/1.5083613}
}

@article{Bittner2020ExcitonInteractions,
  author  = {Bittner, Eric R. and Li, Hao and Piryatinski, Andrei and Srimath Kandada, Ajay Ram and Silva, Carlos},
  title   = {Probing exciton/exciton interactions with entangled photons: Theory},
  journal = {The Journal of Chemical Physics},
  volume  = {152},
  number  = {7},
  pages   = {071101},
  year    = {2020},
  doi     = {10.1063/1.5139197}
}

@article{Li2020LNModulator,
  author  = {Li, Mingxiao and Ling, Jingwei and He, Yang and Javid, Usman A. and Xue, Shixin and Lin, Qiang},
  title   = {Lithium niobate photonic-crystal electro-optic modulator},
  journal = {Nature Communications},
  volume  = {11},
  pages   = {4123},
  year    = {2020},
  doi     = {10.1038/s41467-020-17950-7}
}

@article{Multani2026MillimeterWaveEO,
  author  = {Multani, Kevin K. S. and Herrmann, Jason F. and Nanni, Emilio A. and others},
  title   = {Integrated millimeter-wave cavity electro-optic transduction},
  journal = {Nature Communications},
  volume  = {17},
  pages   = {1166},
  year    = {2026},
  doi     = {10.1038/s41467-025-67932-w}
}

@article{Kuznetsov2026FloquetPolaritons,
  author  = {Kuznetsov, Alexander S. and Carraro-Haddad, Ignacio and Usaj, Gonzalo and Biermann, Klaus and Fainstein, Alejandro and Santos, Paulo V.},
  title   = {Ground-state exciton--polariton condensation via coherent {Floquet} driving},
  journal = {Nature Photonics},
  volume  = {20},
  pages   = {586--591},
  year    = {2026},
  doi     = {10.1038/s41566-026-01855-w}
}

@article{Shalabney2015MolecularResonators,
  author  = {Shalabney, Atef and George, Jino and Hutchison, James A. and Pupillo, Guido and Genet, Cyriaque and Ebbesen, Thomas W.},
  title   = {Coherent coupling of molecular resonators with a microcavity mode},
  journal = {Nature Communications},
  volume  = {6},
  pages   = {5981},
  year    = {2015},
  doi     = {10.1038/ncomms6981}
}

@article{Zhong2015RareEarthNanocavity,
  author  = {Zhong, Tian and Kindem, Jonathan M. and Miyazono, Evan and others},
  title   = {Nanophotonic coherent light--matter interfaces based on rare-earth-doped crystals},
  journal = {Nature Communications},
  volume  = {6},
  pages   = {8206},
  year    = {2015},
  doi     = {10.1038/ncomms9206}
}
\end{document}